\documentclass[conference]{IEEEtran}
\IEEEoverridecommandlockouts
\usepackage{cite}
\usepackage{amsmath,amssymb,amsfonts}
\usepackage{algorithmic}
\usepackage{graphicx}
\usepackage{textcomp}
\usepackage{xcolor}
\usepackage[normalem]{ulem}
\def\BibTeX{{\rm B\kern-.05em{\sc i\kern-.025em b}\kern-.08em
    T\kern-.1667em\lower.7ex\hbox{E}\kern-.125emX}}
\begin{document}

\title{What are the Practices for Secret Management in Software Artifacts?}

\author{\IEEEauthorblockN{Setu Kumar Basak\IEEEauthorrefmark{1},
Lorenzo Neil\IEEEauthorrefmark{2}, Bradley Reaves\IEEEauthorrefmark{3} and
Laurie Williams\IEEEauthorrefmark{4}}
\IEEEauthorblockA{Department of Computer Science,
North Carolina State University, USA\\
Email: \IEEEauthorrefmark{1}sbasak4@ncsu.edu,
\IEEEauthorrefmark{2}lcneil@ncsu.edu,
\IEEEauthorrefmark{3}bgreaves@ncsu.edu,
\IEEEauthorrefmark{4}lawilli3@ncsu.edu}}

\maketitle

\begin{abstract}
Throughout 2021, GitGuardian's monitoring of public GitHub repositories revealed a two-fold increase in the number of secrets (database credentials, API keys, and other credentials) exposed compared to 2020, accumulating more than six million secrets. A systematic derivation of practices for managing secrets can help practitioners in secure development. \textit{The goal of our paper is to aid practitioners in avoiding the exposure of secrets by identifying secret management practices in software artifacts through a systematic derivation of practices disseminated in Internet artifacts}. We conduct a grey literature review of Internet artifacts, such as blog articles and question and answer posts. We identify 24 practices grouped in six categories comprised of developer and organizational practices. Our findings indicate that using local environment variables and external secret management services are the most recommended practices to move secrets out of source code and to securely store secrets. We also observe that using version control system scanning tools and employing short-lived secrets are the most recommended practices to avoid accidentally committing secrets and limit secret exposure, respectively.
\end{abstract}


\section{Introduction} \label{Introduction}
In March 2022, GitGuardian stated that the number of secrets exposed on public GitHub repositories doubled in 2021 when compared to 2020, reaching a total of over six million secrets \cite{GitGuardian}. Software uses external web services for essential functionality. APIs for payment systems, location services, and social networking platform integration, to name a few, are all examples of external web services. To perform authentication across software artifacts as part of system integration, software developers need secrets (database credentials, API keys, and other credentials). During software development, these secrets may need to be shared by developers working on a team, and after deployment may need to be distributed to applications.

In 2019, Meli et al. studied a 13\% snapshot of public GitHub repositories and found over 200K API keys and tokens checked into the repositories \cite{meli2019bad}. Secrets are not only pushed into version control system (VCS) repositories by developers but they are also kept in Android and iOS application packages. One in every 200 Android apps is leaking sensitive information, such as Twitter and AWS API keys \cite{mobile-app-leak}, according to a security research firm that reverse-engineered approximately 16K Android apps. Secrets in software artifacts (CWE-798: Use of Hard-coded Credentials \cite{cwe-798}) have also been identified as a CWE Top 25 Most Dangerous Software Weaknesses \cite{cwe-top-25}. 

The presence of software secrets in VCS repositories necessitates the integration of adequate secret management practices for secure development. However, such integration may be difficult due to the lack of a comprehensive set of practices related to managing secrets. For example, developers seem to query online forums to find the best practices for storing secrets \cite{so-example}. Secret management practices can be derived systematically to help practitioners in limiting the exposure of secrets. In addition, the derived set of practices can be utilized by practitioners as a comparison point for their existing secret management practices.

Analyzing Internet artifacts, such as blog articles and online forum question and answer (Q\&A) posts, is one way to derive secret management practices in software artifacts. In previous studies, the importance of Internet artifacts has also been recognized in determining security practices \cite{IACbestpractice, DevOpsGrey}.

\textit{The goal of our paper is to aid practitioners in avoiding the exposure of secrets by identifying secret management practices in software artifacts through a systematic derivation of practices disseminated in Internet artifacts.}

We answer the following research question: \textbf{RQ: What are the practices used by practitioners for secret management in software artifacts?}

We conducted a grey literature review \cite{greyliteraturedef} and collected 54 Internet artifacts, such as blog articles and Q\&A posts. From the collected Internet artifacts, we conducted a qualitative analysis approach called open coding \cite{opencodinginbook} and determined practices that are specific to secret management in software artifacts.

\textbf{Our contribution} is a set of practices that practitioners can follow to avoid exposure of secrets in software artifacts.

The rest of our paper is structured as follows: The methodology used in our work is described in Section \ref{Methodology}. We provide our findings in Section \ref{Results}. Section \ref{RelatedWork} summarizes previous research findings that are pertinent to our paper. The implications and limitations of our paper are addressed in Section \ref{Discussion} Finally, Section \ref{Conclusion} draws the paper's conclusion.

\section{Methodology} \label{Methodology}
To identify practices used by practitioners for secret management in software artifacts, two authors conduct a grey literature review independently. Grey literature is defined as ``... literature that is not formally published in sources, such as books or journal articles" \cite{greyliteraturedef}. A grey literature review differs from a systematic literature review (SLR) or a systematic mapping study (SMS) as researchers leverage peer-reviewed literature indexed in scholar databases in the case of SLR or SMS. Grey literature review, on the other hand, makes use of non peer-reviewed artifacts such as online videos, blog articles, and Q\&A posts that are available on the Internet \cite{greyliteratureinbook}. Grey literature provides better coverage of emerging research topics \cite{GAROUSI2019101, IACbestpractice}. We are inspired by academics who have analyzed Internet artifacts to determine security practices that can be used in the software development process \cite{IACbestpractice, DevOpsGrey}. We hypothesize that by collecting and analyzing Internet artifacts systematically, we can find practices for secret management in software artifacts.

\begin{figure}
    \includegraphics[width=\columnwidth]{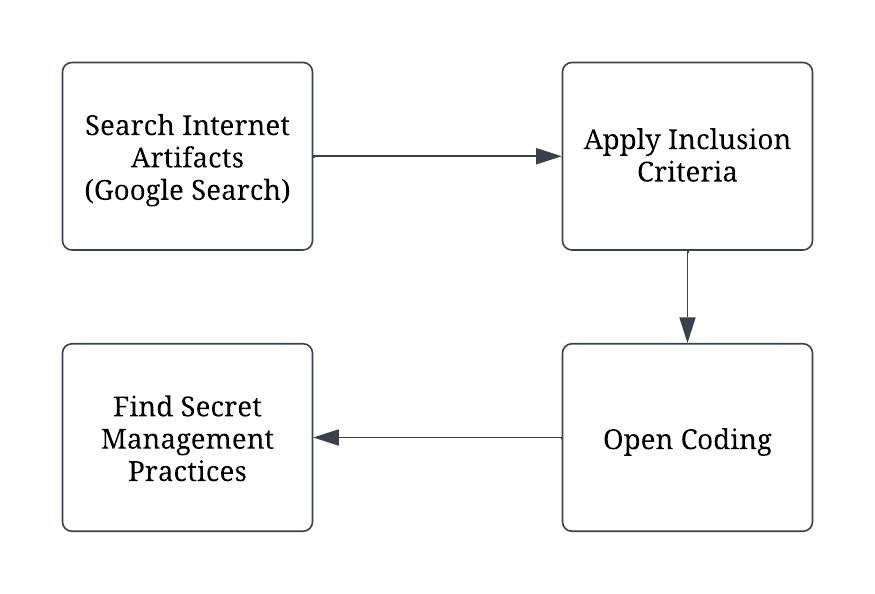}
    \caption{An overview of our grey literature review methodology.}
    \label{fig:greyliteraturesearch}  
\end{figure}

Figure \ref{fig:greyliteraturesearch} shows a summary of our grey literature review methodology. The following is a breakdown of each stage in our methodology.

\subsection{Search Internet Artifacts}
Using a set of search strings, we collect Internet artifacts. As our research study focuses on the practices for managing secrets in software artifacts, we begin with the search string ``practice for managing secrets in source code". Next, we choose the top 100 results determined by Google search engine's page rank algorithm as a search stopping criteria \cite{GAROUSI2019101} and collect the results. We observe practitioners referring to ``secrets" as ``credentials", ``passwords" and ``sensitive information" based on a manual examination of the 100 search results. Based on the observations above, we include these keywords as part of the search construction process and conduct our search procedure using four search strings, which are stated below:

\begin{itemize}
    \item practice for managing secrets in source code
    \item practice for managing credentials in source code
    \item practice for managing passwords in source code
    \item practice for managing sensitive information in source code
\end{itemize}

Altogether, we collect 400 Internet artifacts, 100 for each of the four search strings. To avoid a conflict with the authors' browsing history, we search in incognito mode of the Google Chrome browser.

\subsection{Apply Inclusion Criteria}
To find Internet artifacts relevant to our research study, we use the following inclusion criteria: 

\begin{itemize}
    \item The artifact is not a duplicate of another artifact;
    \item The artifact is available for reading;
    \item The artifact is written in English; and
    \item The artifact discusses at least one practice for secret management in software artifacts
\end{itemize}

\begin{figure}
    \centering
    \includegraphics[]{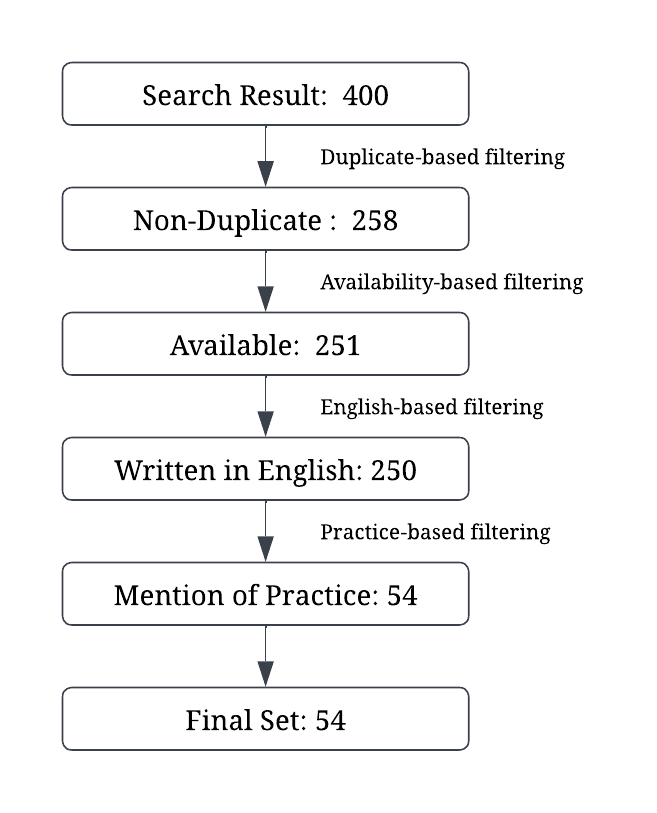}
    \caption{Application of inclusion criteria on our grey literature dataset to collect the set of 54 Internet artifacts that we use in our study. Grey literature dataset is available online \cite{dataset}.}
    \label{fig:greyliteratureresult}  
\end{figure}

We determine 54 Internet artifacts after applying the inclusion criteria. Three Q\&A posts and 51 blog articles comprise our collection of 54 Internet artifacts. Figure \ref{fig:greyliteratureresult} depicts a detailed breakdown of our filtering procedure.

\subsection{Find Secret Management Practices} We apply open coding \cite{opencodinginbook} to our collected grey literature artifacts. Open coding is a qualitative analysis technique that can reveal the underlying theme from unstructured textual information \cite{opencodinginbook}. Open coding is widely utilized to identify patterns from Internet artifacts \cite{IACbestpractice, akondtestpractice}. The first and second authors review each Internet artifact and extract the stated practices as part of the open coding process. After the first and second authors finish their open coding individually, the identified practices are cross-checked by both authors. We use a negotiated agreement \cite{negotiatedagreement} to resolve the disagreed-upon practices. Negotiated agreement is an approach to discuss the disagreements among the raters in an effort to resolve disagreements when two or more raters code the same artifacts \cite{negotiatedagreement}. We resolve disagreements either by discarding practices that are not suitable for managing secrets or combining similar practices into one practice. We group each identified practice into a category that solves a specific issue of secret management, such as developer practices for avoiding accidental secrets commit or organizational practices to enforce policies for secrets protection.  

\section{Results} \label{Results}
Based on our findings, we identified 24 practices classified into six categories. In the following subsections, we provide the details of the identified practices of each of the six categories and the number of Internet artifacts that discuss the practices. For example, the `Practices for Keeping Secrets Out of Source Code (OSC)' category has four practices, and 18 of the 54 Internet artifacts mention the `OSC-1: Use Local Environment Variables' practice.

\subsection{\textbf{Practices for Keeping Secrets Out of Source Code (OSC)}}
Beyond allowing developers to set permissions on their accounts, VCS does not ensure the security of secrets to remain in a secured and controlled environment. Practitioners recommend the below four practices to keep secrets out of source code or VCS repositories.
 
\subsubsection{\uline{OSC-1: Use Local Environment Variables (18)}}

Local environment variables, which are dynamic objects defined outside of the application and used to avoid storing secrets in VCS or configuration (config) files, are recommended by practitioners. The Twelve-Factor App methodology \cite{12-factor-app} is a set of 12 language-agnostic guidelines for building software-as-a-service applications with portability and resilience. The third factor, \textit{Config principle} of the Twelve-Factor App methodology, also states that config information should be kept as environment variables and injected into the application at runtime \cite{12-factor-app}. Libraries, such as \texttt{dotenv} \cite{dotenv}, can be used to load variables into the running process. Without modifying any code, environment variables can easily be changed between released versions. On the other hand, practitioners also advise avoiding local environment variables for client-side applications as secrets can be extracted using the browser dev tools \cite{local-env-client}.

\subsubsection{\uline{OSC-2: Move Secrets to Configuration File (15)}} 

Practitioners recommend moving secrets to external config files such as \texttt{web.config} and \texttt{config.yaml} files. Config files are environment-specific and can be updated at any time without redeploying the application, as the lifecycle is independent of the application. Instead of checking the original config file into VCS, developers are suggested to add a template config file. The template config files such as \texttt{database.sample.yaml} file of Ruby on Rails and \texttt{web.default.config} file of ASP.NET will contain minimum configurations with dummy values which developers will replace in their development environment. Using template config files reduces the chance of secrets being checked into VCS, thus preventing potential secret exposure. 

\subsubsection{\uline{OSC-3: Ignore Sensitive Files (11)}}

Practitioners recommend to avoid committing sensitive files, such as .env and .config files, into VCS repositories. Even a \texttt{.DS\_Store} (a hidden file present in every folder on an OS X system) can leak the names of the files and folders present on a web server. A search on GitHub for \texttt{.DS\_Store} returns more than 800K results \cite{internetwachedsstore}. To avoid committing sensitive files, all repositories should include a \texttt{.gitignore} file. GitHub has published a collection of useful \texttt{.gitignore} templates for different technologies \cite{github-gitignore-template}.

\subsubsection{\uline{OSC-4: Add Server-Side Implementation for Client-Side Applications (2)}}

Secrets present in client-side applications, such as Javascript and Android applications, can be exposed by the developer console or by decompiling the binary files (APK or iPA files) \cite{mobile-app-leak}. To avoid keeping secrets in client-side applications for fetching data from different web services, practitioners recommend implementing the web service functionality on the server-side. The server will use the appropriate secrets and fetch data for the client-side, thus removing the necessity to keep secrets in client-side applications.

\subsection{\textbf{Practices for Securely Storing Secrets (SSC)}}
Developers can store secrets insecurely in source code or VCS repositories. Practitioners recommend the following three practices to store secrets securely.

\subsubsection{\uline{SSC-1: Use External Secret Management Systems (28)}}
As emails can be forwarded and messaging applications can be hacked, practitioners recommend to avoid sending secrets through emails or any messaging applications, such as Microsoft Teams or Slack. Only one compromised account is enough to expose sensitive data. When secrets are exchanged through internal networks, bad actors can use secrets to migrate laterally between services. Instead, practitioners recommend to use external secret management systems, such as \texttt{HashiCorp Vault} \cite{hashicorp-vault}, \texttt{AWS KMS} \cite{aws-kms} and \texttt{Knox} \cite{knox}. These hardware security modules can safely store secrets with tightly-controlled access. Developers can be assigned to groups based on the teams they work on, and secrets can be shared with the groups by granting proper permissions. If any developer switches teams or leaves the company, the secrets used by the developer can easily be invalidated using external secret management systems. The ability to set up dynamic secrets, lease-based secret management (limiting access for a fixed period before automatic revocation), and audit trails, which allow administrators to check for any breaches, are other important features. The ability to rotate secrets over time by giving specific states for secret versions is a unique feature of \texttt{Knox} that is not found in other systems. A secret version can be tagged as `primary' to denote that the secret is the current recommended, `active' to denote that the secret is still usable, or `inactive' to denote that the secret is disabled. Administrators can use this mechanism to roll secrets across machines without impacting the service. External secret management systems minimize human involvement in creating, distributing, and maintaining secrets. Practitioners have recommended SSC-1 practice the most among all the practices for managing secrets, though a significant investment of time and money is needed.

\subsubsection{\uline{SSC-2: Store Encrypted Secrets (14)}}

Practitioners recommend avoiding Base64 encoding of secrets as encoded secrets can be decoded easily. Instead, for a project having a single developer or a small team, practitioners suggest encrypting secrets-containing files in VCS. Several tools such as \texttt{git-crypt} \cite{git-crypt} and \texttt{git-secret} \cite{git-secret} which use GPG to encrypt content are available for encrypting sensitive files containing secrets. Technologies, such as Ruby on Rails, starting with Version 5.1, have included built-in capabilities to encrypt secrets with VCS \cite{ruby-on-rails-secrets}. Though developers have to manage encryption keys securely (keep out of VCS) and no role-based access control of secrets is present, the benefit of using encryption tools is that the implementation does not need additional infrastructure.

\subsubsection{\uline{SSC-3: Private Repositories Are Not Safe (2)}}

One practitioner stated: ``\textit{A secret in a private repo is like a password written on a \$20 bill, you might trust the person you gave it to, but that bill can end up in hundreds of peoples hands as a part of multiple transactions and within multiple cash registers}" \cite{bestpracgitguardian}. Since repositories can be forked into new projects and cloned onto new machines, secrets present in the history of a repository will be propagated to the forked and cloned repositories. Only one compromised developer account or a misconfiguration will be enough to get access to all secrets present in private repositories. For example, in 2021, a repository misconfiguration of setting the default username and password combination of \texttt{admin/admin} resulted in Nissan source code being exposed online \cite{nissan-leak}.

\subsection{\textbf{Practices to Limit Secrets Exposure (LSE)}}
Practitioners recommend below four practices to limit the exposure of secrets.

\subsubsection{\uline{LSE-1: Use Short-lived Secrets (15)}}

Short-lived secrets, according to practitioners, prevent previously-undetected data breaches from becoming a threat by terminating access even if the breach is not identified. If a validity period cannot be assigned to secrets, practitioners advise revoking and redistributing the secrets periodically. Practitioners also suggest rotating and redistributing the secrets correctly to avoid any failure. For example, in 2021, Microsoft Azure experienced a 14-hour downtime due to an error in secret rotation used for authentication \cite{azure-outage}.

\subsubsection{\uline{LSE-2: Restrict API Access and Permissions (8)}}

Because attackers frequently use secrets within their scope, detecting when they are doing so maliciously may be challenging. Practitioners suggest that damage and lateral movement can be limited by restricting access and permissions to secrets. For example, a leaked AWS S3 key, which had the permission to spin up AWS EC2 instances, resulted in a \$6000 bill overnight as an attacker spun up 140 instances \cite{aws-6000-leak}. IP white-listing adds another degree of protection against attackers who try to exploit APIs maliciously. For example, GitHub IP white-listing \cite{github-white-list} can be employed to restrict insecure sources from accessing the repositories. External service users can set API key usage restrictions by making API keys accessible from specific URLs. The key will be useless to the attacker if the attacker cannot invoke the service from the allowed URLs. A daily limit on API key usage should also be set to avoid bill spikes. 

\subsubsection{\uline{LSE-3: Revoke Secrets and Sanitize VCS History (6)}}

Secrets will not be removed entirely by removing them in another commit, as secrets will remain in the VCS history. Practitioners advise sanitizing VCS history in two steps. The first step is to revoke the secrets present in the code. The second step is to purge and rewrite the VCS history using tools such as \texttt{git-filter-branch} \cite{git-filter-branch}, \texttt{BFG repo cleaner} \cite{bfg-repo-cleaner}, or \texttt{git-filter-repo} \cite{git-filter-repo}. GitHub documentation \cite{github-rewrite-doc} suggests using \texttt{BFG repo cleaner} instead of other tools. To avoid anomalies, the best practice is to close all pull requests before scanning VCS history using the tools. GitHub suggests contacting them with the repository name to clear the secrets from their cache and advised to tell the project collaborators to do \texttt{git rebase} instead of \texttt{git merge} as merge can introduce some of the tainted history \cite{github-rewrite-doc}.

\subsubsection{\uline{LSE-4: Audit All Code Uploaded to VCS and Review VCS Audit Logs for Suspicious Activity (4)}}

Practitioners recommend auditing all code uploaded to VCS on a regular basis. For example, legacy code may be used as part of an organization's new software. The problem with integrating legacy code is that what was once secure might not be anymore, as secrets may be present in legacy code. Therefore, auditing any code uploaded to VCS will be advantageous for the software's long-term integrity, even if the procedure is time-consuming. The administrator of an organization can also review the activities of other team members using the audit log feature of VCS. Suspicious activities can be flagged and tracked by constructing a trace profile based on the user's activity, the action's location, and the time of the event. For example, GitLab \cite{gitlab-audit-log} provide the archive of audit logs where the admin can search for events between any period or any specific user action.

\subsection{\textbf{Practices for Avoiding Accidental Secrets Commit (ASC)}} 

Developers can accidentally push secrets into VCS repositories. Practitioners recommend the below three practices to avoid accidental committing of secrets.

\subsubsection{\uline{ASC-1: Use VCS Scan Tools (16)}}

Though code reviews can detect logical flaws and maintain coding practices, practitioners do not recommend relying on code reviews to detect secrets. If secrets are added in one commit and removed in another commit, the net difference in code changes will be zero. The reviewer only sees the net difference, but secrets will remain in the VCS history, thus allowing an attacker to find secrets from the prior revisions. Practitioners recommend running VCS scan tools such as \texttt{TruffleHog} \cite{trufflehog}, \texttt{Gitrob} \cite{gitrob} and \texttt{git-all-secrets} \cite{git-all-secrets} in the VCS repository to find out the presence of secrets. Via a \texttt{pre-commit} hook, VCS scan tools can reject any commit containing secrets that manual searches and reviews will miss. VCS scan tools can also find secrets buried in logs and histories. VCS scan tools are also recommended to use with continuous integration or continuous deployment (CI/CD) pipelines to actively break build/deploy when secrets are found in source code. Practitioners also mention that VCS scan tools will return a lot of false positives, which developers will need to filter manually \cite{rahman2022secret}.

\subsubsection{\uline{ASC-2: Add Files to the Staging Area Explicitly (5)}}

One simple practice suggested by practitioners to avoid exposing secrets accidentally is to add files explicitly in the VCS staging area. Developers should avoid using wildcard commands (\texttt{git add -A}, \texttt{git add .} and \texttt{git add *}) when adding files to have full control over what is committed. One practitioner also suggested: ``\textit{Committing early and committing often will not only help navigate file history and break up otherwise large tasks, in addition it will reduce the temptation to use wildcard commands.}" \cite{bestpracgitguardian}.

\subsubsection{\uline{ASC-3: Use VCS Hooks to Check Files Prior to Committing (5)}}

To prevent secrets from pushing into VCS repositories, practitioners advise implementing VCS hooks \cite{git-hook} which allow scripts to be executed before or after a specific action in the VCS repository. The \texttt{pre-commit} and \texttt{post-commit} hooks can be used to filter and smudge secrets before commit or after pull, respectively \cite{git-smudge-clean}. Each contributor to the VCS repository needs to set up VCS hooks individually. According to practitioners, VCS hook scripts need extra effort to write properly since putting all of the secret behavior in the script is challenging.

\subsection{\textbf{Practices for Managing Secrets in Deployment (MSD)}}
Developers can expose secrets during deployment.  Practitioners recommend the following four practices to manage secrets in deployment securely.

\subsubsection{\uline{MSD-1: Use Secret Variables in CI/CD (6)}}

Practitioners recommend removing hard-coded secrets from CI/CD scripts, and use the secret variables of the build/deploy systems, such as \texttt{Heroku} \cite{heroku-config-vars} and \texttt{Azure Pipeline} \cite{azure-pipeline-vars}. VCS such as GitHub \cite{github-secrets} and GitLab \cite{gitlab-ci-variables} have also secret variables which can be used in the CI/CD pipeline. The secret variables are set as environment variables in the deployment environment and hidden from any logs. Practitioners also suggest keeping secret variables settings disabled for pull requests to avoid inadvertently passing secrets during builds for pull requests.

\subsubsection{\uline{MSD-2: Use Configuration Management Systems (4)}}

The configurations of different machines are coordinated by Configuration Management System (CMS) tools from a centralized location. Practitioners recommend using secret management systems supported by CMS tools, such as \texttt{Chef-Vault} \cite{chef-vault} and \texttt{Ansible-Vault} \cite{ansible-vault}. Using shared secrets, these CMS tools keep secrets out of revision history and from other machines. Secrets can be distributed to specific machines using the same mechanism which ensures each machine receives the correct configuration.

\subsubsection{\uline{MSD-3: Use Different Secrets for Each Environment (3)}} 

Practitioners recommend to avoid using the same secrets for multiple environments so that exposure to the secrets of one environment does not compromise other environments. The secrets of production environments should be different from development or pre-production environments. Practitioners also recommend keeping production environment secrets limited to a small set of owners to avoid the risk of failure.

\subsubsection{\uline{MSD-4: Keep Dot Files out of Root Directory (2)}}

During deployment, practitioners recommend keeping dot files, such as \texttt{.git}, \texttt{.gitignore} and \texttt{.env} files, out of the root directory. Proper access restrictions should be applied to dot files on production servers to avoid exposing secrets \cite{dot-git-expose}. If the \texttt{.git} folder is not kept out of the root directory, then the whole history of committed changes will be exposed to the attacker. Previous research \cite{meli2019bad} has also found secrets in the \texttt{.gitignore} file despite the \texttt{.gitignore} file is designed to restrict unintended source files committing into VCS.

\subsection{\textbf{Organizational Practices to Enforce Policies for Secrets Protection (OEP)}}
Organizations can adopt general practices to enforce policies in VCS for developers. These general practices can minimize vulnerabilities which in turn helps in avoiding exposure of secrets. Practitioners recommend the below six practices for enforcing policies.

\subsubsection{\uline{OEP-1: Tightly Manage Developer Permissions (6)}}
According to practitioners, organizations should follow the principle of least privilege. Organizations should not give developers more permissions than the required scope, such as changing repository visibility and adding external contributors. If the repository contains secrets, the more developers who have permission to change the visibility of the repository, the higher the risk of failure. For example, GitHub has organization-level settings to restrict the ability to change the visibility of the repository to anyone with admin access or organization owners \cite{github-repo-visibility}. One practitioner stated: ``\textit{The higher the turnover of external contributors, the higher the security risks}" \cite{spectralops-practices}. External contributors can be strictly managed to reduce the number of redundant developers and their access to the repositories. Limiting access and permission-granting privileges to organization owners is one way to handle external contributors. For example, GitHub has organization-level settings for allowing the permission to add external contributors to the organization owners only \cite{github-ext-contributor}.

\subsubsection{\uline{OEP-2: Enforce Two-Factor Authentication (6)}}

To prevent source code leakage via insecure developer accounts, practitioners recommend enforcing two-factor authentication (2FA). When logging into VCS, such as GitLab and GitHub, 2FA provides an added layer of security which can be enforced through organization-level settings \cite{github-2fa, gitlab-2fa}. 

\subsubsection{\uline{OEP-3: Require Commit Signing (3)}}

A malicious user can push exploitable code into VCS  by pretending to be someone else and remain untraceable by altering the username and email address in the \texttt{git config}. For verification and traceability of code merge, commit signing \cite{github-commit-sign}, which is a cryptographic code-signing technique, can be used. Commit signing is done through GPG, and the signed commit gets a `verified' badge. A malicious user's commit can easily be traced as the commit will not have a `verified' badge.

\subsubsection{\uline{OEP-4: Add a security.md File (3)}}

Practitioners recommend adding a \texttt{security.md} file \cite{github-security-md} in a VCS repository to officially document security-related processes and procedures, such as token accessibility, authentication requirements, and vulnerability reporting. The \texttt{security.md} file can serve as a helpful reference for developers as well as a centralized space for the organization's security expectations.

\subsubsection{\uline{OEP-5: Implement Single Sign-On (2)}}

SAML single sign-on (SSO) \cite{github-saml} is a VCS feature practitioners recommend. Access to VCS resources, such as specific repositories and pull requests, can be managed by explicitly providing permissions to resources by leveraging SAML SSO. SAML SSO also allows to set up of approved identity providers, which enables the organization to force the developers to sign in using the organization's accounts instead of privately-owned VCS accounts \cite{github-saml}.

\subsubsection{\uline{OEP-6: Disable Forking (1)}}

A Git feature called forking allows a developer to duplicate a repository and is useful for testing and sandboxing. However, practitioners recommend to disable forking. A fork can reveal secrets to the public though the repository is originally private. With each fork, the risk grows exponentially, resulting in a chain of security vulnerabilities. For example, GitHub has organization-level settings to disallow forking of repositories \cite{github-fork-policy}.

\section{Related Work} \label{RelatedWork}
Prior work has found that root causes for widespread secret leakage were insecure developer practices, such as embedding hard-coded credentials \cite{MedicalDataLeak,GithubLeaks}, organizational issues influencing software security vulnerabilities \cite{assal2019think,xie2011programmers,nadi2016jumping, rahman2022secret}, and compromising security for functionality when managing software dependencies \cite{pashchenko2020qualitative}. Researchers have looked into instances of such insecure developer practices within open-source projects \cite{meli2019bad,7180102,rahman2019share}. Researchers have discovered hard-coded secrets as a very common development practice which has resulted in thousands of repositories on open-source coding platforms, such as GitHub and OpenStack, leaking hard-coded secrets \cite{meli2019bad,rahman2019share, rahman2021different}. Within Infrastructure as Code (IaC) scripts, Rahman et al. \cite{rahman2019seven,rahman2021different,rahman2021security} looked for security smells, which are repeating coding patterns that indicate a security flaw. Results from Rahman's work conclude that developers commonly commit multiple security smells within their development practices, and hard-coded credential is the most occurring smell. In another study, Rahman et al. \cite{IACbestpractice} analyzed Internet artifacts and identified 12 secret management practices related to IaC. The mentioned practices for IaC, such as separation of vaults from VCS and secret rotation, relate to the `Ignore Sensitive Files (OSC-3)' and `Use Short-lived Secrets (LSE-1)' practices, respectively. We take motivation from the aforementioned studies and concentrate our research efforts on finding secret management practices for software artifacts in general, instead of specific technology such as IaC.

\section{Discussion} \label{Discussion}
In this section, we discuss the implications and limitations of our paper.\\
\textbf{\uline{Implication for Practitioners:}}  Our derivation indicates how secrets should be kept out of source code or VCS to avoid being exposed. Insufficient application of practices related to secret management can result in unfavorable outcomes. The identified practices can help practitioners manage secrets in software artifacts and act as a comparison point for practitioners with their existing secret management practices.
\\\textbf{\uline{Implications for Researchers and Tool Developers:}} Further research can be conducted in the field of secret management by leveraging our findings. For example, researchers can look into how many of the practices specified in Section \ref{Results} are followed in commercial and open-source software. We hypothesize that systematic follow of secret management practices is not prevalent based on the presence of secrets in software artifacts. If empirical studies support our hypothesis, researchers can look into the contributing factors to ignoring secret management practices. Our findings will also assist tool developers in determining whether to create new tools or improve existing ones to help practitioners manage secrets more effectively. The false-positive rate of VCS scan tools, for example, can be improved.
\\\textbf{\uline{Limitations:}} Since the identified practices are bound to the Internet artifacts which we analyzed in our study, our findings are subject to external validity and may not generalize to another collection of Internet artifacts. We account for the constraint by employing four search strings to collect and filter Internet artifacts in a systematic way. Manual analysis may induce bias while identifying practices. We account for this bias by adding a second rater who identified practices from the same set of Internet artifacts independently. Finally, the sets of practices of each rater are cross-checked to mitigate bias.

\section{Conclusion} \label{Conclusion}
In addition to users, software also relies heavily on the use of secrets for authentication and authorization, and the exposure of secrets is increasing each day. The presence of secrets in software artifacts substantiates the practitioner's lack of knowledge on securely managing secrets. A set of secret management practices can help practitioners avoid exposing secrets in software artifacts. To identify practices for secret management in software artifacts, we conducted a grey literature review of 54 Internet artifacts. Our analysis identified 24 practices grouped into six categories and comprised of developer and organizational practices. According to our findings, the most recommended practices for moving secrets out of source code and securely storing secrets include using local environment variables (OSC-1) and external secret management services (SSC-1). We also observed that using VCS scan tools (ASC-1) and employing short-lived secrets (LSE-1) are the most recommended practices to avoid accidentally committing secrets and limit secrets exposure. Our findings can also be beneficial for researchers and tool developers who can investigate how the secret management process can be enhanced to facilitate secure development.

\section*{Acknowledgment}

This work was supported by the National Science Foundation grant 2055554.  The authors would also like to thank the North Carolina State University Realsearch research group for their valuable input on this paper.

\bibliographystyle{IEEEtran}
\bibliography{bibliography}

\end{document}